\documentclass[aps,prl,reprint,superscriptaddress]{revtex4-2}
\usepackage{graphicx,subfigure,amsmath}
\usepackage{dcolumn}
\usepackage{bm}
\usepackage{multirow}
\usepackage{ulem}
\usepackage{color}

\begin{document}

\title{Prospects for 10$^{-18}$ Instability Laser Referenced on Thermal Atomic Ensembles}

\author{Haosen Shang} 
\affiliation{State Key Laboratory of Advanced Optical Communication Systems and Networks, Department of Electronics, Peking University, Beijing 100871, China}
\author{Duo Pan} 
\email{panduo@pku.edu.cn}
\affiliation{State Key Laboratory of Advanced Optical Communication Systems and Networks, Department of Electronics, Peking University, Beijing 100871, China}
\author{Xiaogang Zhang}
\affiliation{State Key Laboratory of Advanced Optical Communication Systems and Networks, Department of Electronics, Peking University, Beijing 100871, China}
\author{Xiaobo Xue}
\affiliation{Science and Technology on Metrology and Calibration Laboratory, Beijing Institute of Radio Metrology and Measurement, Beijing 100854, China}
\author{Tiantian Shi}
\affiliation{State Key Laboratory of Advanced Optical Communication Systems and Networks, Department of Electronics, Peking University, Beijing 100871, China}
\author{Jingbiao Chen}
\affiliation{State Key Laboratory of Advanced Optical Communication Systems and Networks, Department of Electronics, Peking University, Beijing 100871, China}

\begin{abstract}
A thermal atomic ensemble-based laser source with superior frequency stability is proposed that relies on the accumulated contributions from an abundance of nonzero-transverse-velocity atomic ensembles. Compared with the traditional case in which only atoms with near-zero transverse velocities are utilized, the amplitude of the optical Ramsey fringes for a thermal calcium beam can be dramatically enhanced by three orders of magnitude or more, thus, the signal-to-noise ratio can be improved 33-fold. Based on the recent results of atomic interferometry-based laser stabilization, a quantum projection noise-limited frequency instability less than 2$\times 10^{-17}/\sqrt{\tau}$ is feasible. Such an ultrastable laser has promising applications in diverse areas, including metrology and astronomy.
\end{abstract}

\date{\today}

\maketitle

\begin{figure}[t]
\begin{center}
\includegraphics[width=8.5 cm,height=6.3 cm]{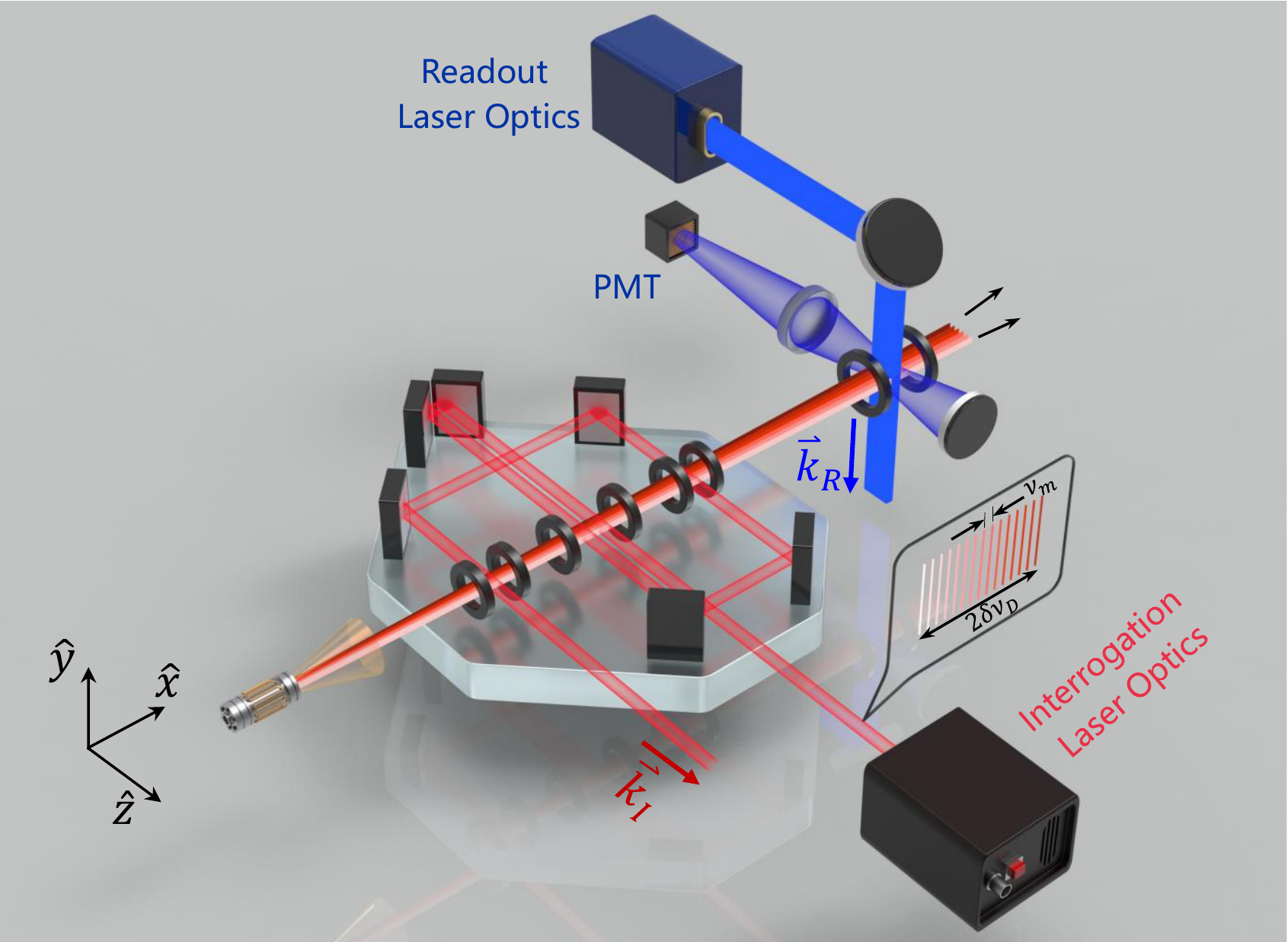}
\end{center}
\caption{Experimental configurations of the ultrastable laser source referenced on thermal Ca beam. Both the optical local oscillator (OLO) laser and readout laser are pre-stabilized for intensity and frequency. The OLO laser is phase-modulated with a frequency of $\nu_{m}$ to be a multi-frequency laser (interrogation laser) by a series of cascade phase modulators. The spectrum of the interrogation laser with a span of $2\delta \nu_{D}$ is shown in the black frame. The gradual colors in the atomic beam denote the atoms with different Doppler shifts. PMT, photomultiplier tube. $\vec{k}_{I}$ and $\vec{k}_{R}$ denote the wavevector of the interrogation laser and readout laser, respectively. Magnetic fields in the interrogation and readout zones are produced by four pairs of black coils.}
\label{F1}
\end{figure}

\begin{figure}[t]
\centering
\subfigure{
\includegraphics[width=8cm,height=5.9cm]{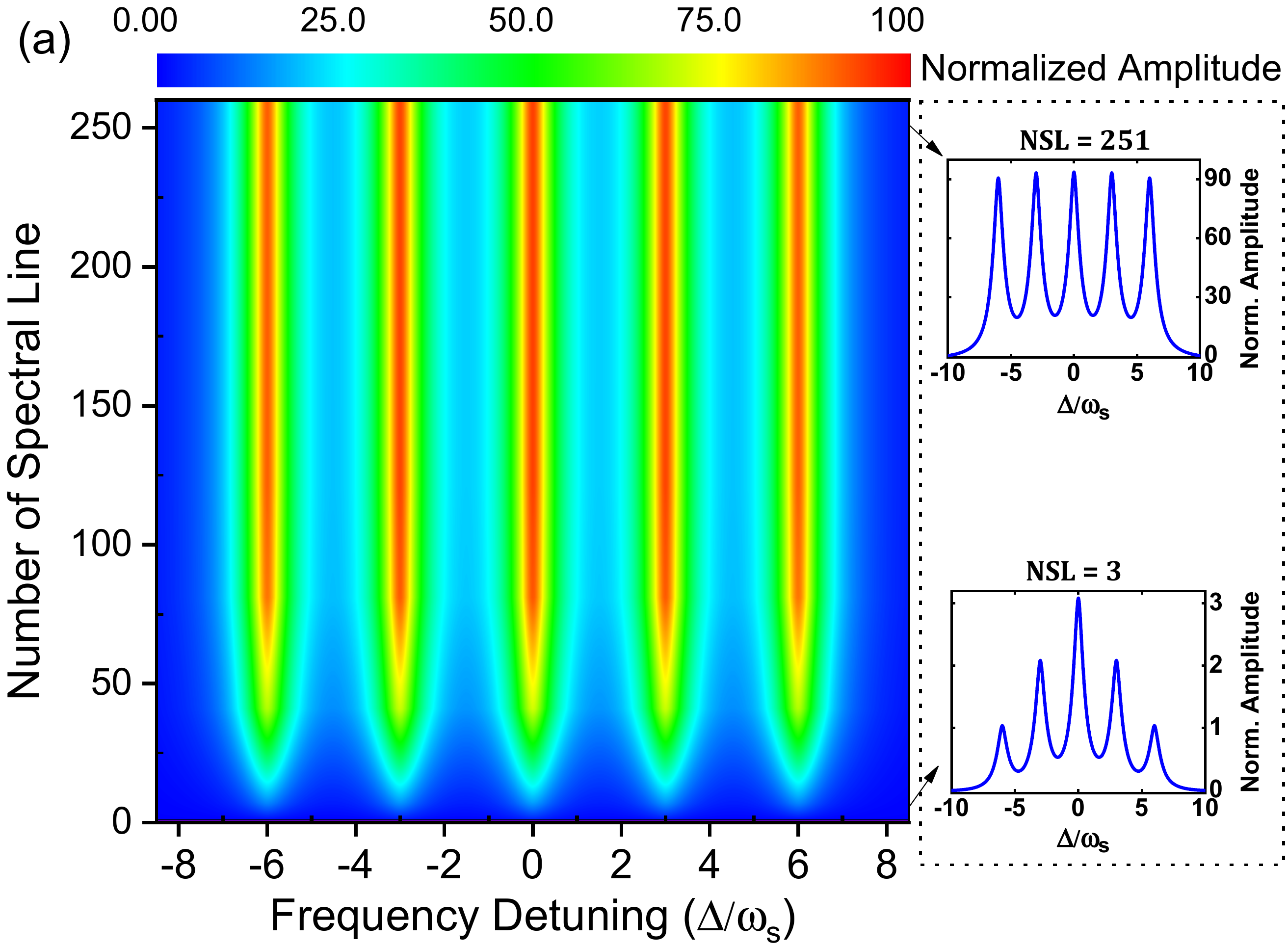}}
\subfigure{
\includegraphics[width=8.1cm,height=2.8cm]{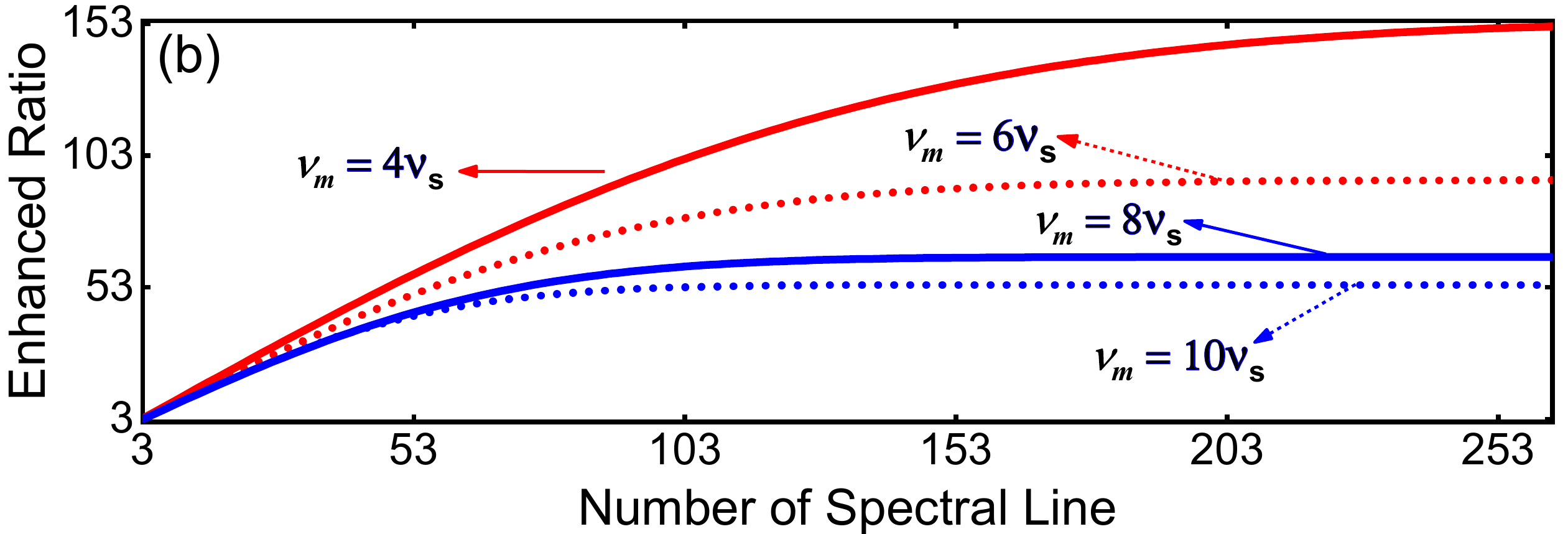}}
\caption{Characteristics of velocity-grating Rabi spectroscopy. (a) Lineshape varied with the $\Delta$ and the NSLs. The upper-right and lower-right insets are the normalized spectroscopy when NSLs=251 and NSLs=3, respectively. Only five central dips are considered here. (b) $\Delta=0$, amplitude-enhanced ratio $\mathcal{R}$ of the spectroscopy vs the NSLs and $\nu_{m}$.} 
\label{F2}
\end{figure}

\begin{figure}[t]
\begin{center}
\includegraphics[width=8.5 cm,height=2.7 cm]{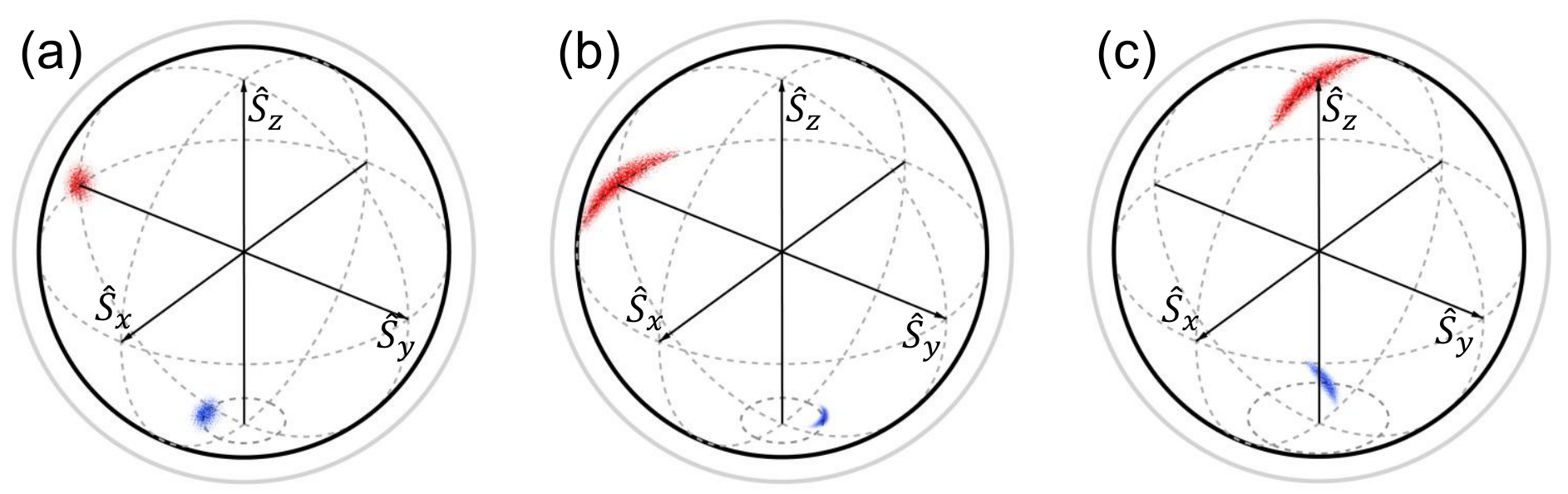}
\end{center}
\caption{Evolution of the pseudo-spin vector in the Bloch sphere for atoms with large transverse velocities at the case of single-frequency laser (blue points) and multi-frequency laser (red points). (a) after the first interrogation field. (b) after free evolution. (c) after the second interrogation field. }
\label{F3}
\end{figure}

Ultrastable lasers with outstanding frequency stability have undergone significant progress in fundamental physics research and advanced technological applications. The latest developments, e.g., gravitational wave detection \cite{Abbott2016} and ultralow phase noise microwave sources \cite{Fortier2011, Xie2017}, reveal its great potential to break through the boundaries of traditional methods. One inspiring application is the optical clock \cite{Ludlow2015, Huang2016}, which enables the frequency measurement to reach an unprecedented level of precision, i.e., 10$^{-19}$ \cite{McGrew2018, Oelker2019}. Such an ability of optical clocks promotes fields in the generation of timescales defined in the optical regime \cite{Milner2019, Nakamura2020}, tests of general relativity \cite{Ashby2018, Takamoto2020}, and monitoring of geopotential \cite{Lisdat2016, Takano2016} based on clock networks \cite{Riehle2017}. 
\\\indent
Recent efforts \cite{Matei2017, Zhang2017, Robinson2019} have promoted the performance of frequency-stabilized lasers referenced on the cryogenic Si cavity to reach a thermal noise-limited instability of $4\times10^{-17}$ \cite{Oelker2019}. More stable lasers require lower thermal cavity noise and subsequently more technical challenges \cite{Matei2017, Zhang2017}, thus motivating the research on alternative approaches, including spectral hole burning \cite{Cook2015} and active optical clocks (superradiant lasers) \cite{Chen2009, Meiser2009, Norcia2018, Schaffer2019}. Considering the remarkable signal-to-noise ratio (\textit{S}/\textit{N}) contributed from numerous thermal atoms, a thermal atomic ensemble with narrowband resonance (e.g., the narrow intercombination transition of alkaline-earth atoms) is an eminent choice for laser-frequency stabilization. Unfortunately, the low utilization of thermal atoms due to Doppler effects limits the huge potential of thermal-ensemble-based laser stabilization, which can be properly solved by the method proposed here.
\\\indent
Because of the pivotal application of spatially separated electronic-shelving detection \cite{Huang2006}, the thermal Ca beam atomic interferometry-based laser stabilization initially showed impactful frequency stability \cite{McFerran2010}. Subsequently, the frequency instability of the laser was further reduced to the level of $10^{-15}$ even with Rabi excitation \cite{Shang2017}. Recently, such an approach impressively demonstrated a frequency instability of 6$\times10^{-16}/\sqrt{\tau}$ in short timescales and presented superior long-term stability and accuracy over the optical cavity \cite{Judith2019, Judith2019T}. However, the Ramsey or Rabi excitation in thermal beams generally encounters the challenges of taking full advantage of atoms with nonzero transverse velocities \cite{McFerran2010, Shang2017, Judith2019, Judith2019T}. Undoubtedly, only a fraction of atoms ($<1\%$) among the large transverse velocity distribution contribute to the \textit{S}/\textit{N}, wasting nearly a trillion atomic references. 
\\\indent
In this paper, a thermal ensemble-based ultrastable laser source is put forward based on the proposed velocity-grating atomic spectroscopy. In this spectroscopy, a series of atoms with equal transverse velocity intervals can be interrogated by a multi-frequency laser with thousands of narrow-linewidth sidebands, thereby tremendously improving the amplitude of the spectroscopy and the \textit{S}/\textit{N}. All the atoms that can be interrogated comprise a grating-like distribution in the transverse velocity domain. Specifically, combining the results reported in \cite{Judith2019, Judith2019T}, a quantum projection noise (QPN)-limited frequency instability less than 2$\times 10^{-17}/\sqrt{\tau}$ can be obtained, which is comparable with the best-reported optical clocks \cite{Oelker2019} in short timescales. Such a laser source with superb frequency stability opens new applications in metrology and astronomy. Although Ca is used as an example here, the scheme is universal for different thermal ensembles.
\\\indent
The experimental scheme of the ultrastable laser source is illustrated in Fig. \ref{F1}, wherein four $\pi$/2 lasers repeatedly interrogate the atomic beam at the case of Ramsey excitation. When only a pair of counter-propagated $\pi$ lasers perpendicularly interact with the atomic beam, the configuration turns out to be the Rabi excitation. The quantum absorber used here is the neutral Ca atom, possessing the $^{1}S_{0}$ -- $^{3}P_{1}$ transition (657 nm, $\sim$ 400 Hz linewidth) insensitive to external field perturbations. For the directional atoms effusing from the oven heated to 625 $^{\circ}$C, the unavoidable transverse divergence would lead to a Doppler shift relevant to velocity along the \textit{$\hat{z}$} direction. One typical method of suppressing the adverse Doppler effect, namely a slit placed behind the oven nozzle, is usually implemented at the cost of the great reduction of quantum absorbers. But in the configuration proposed here, the downstream slit is removed, then a Doppler width can be typically up to 100 MHz \cite{Trager1980, Riehle1988}.
\\\indent
An optical local oscillator (OLO, with an expected Hz-level linewidth) is obtained after the 657 nm laser pre-stabilized by an ultralow-expansion (ULE) cavity \cite{Judith2019, Judith2019T} passing through an AOM, which is used to make up the frequency difference between the cavity and atoms. Before interacting with the atoms, the OLO is phase-modulated to be a multi-frequency interrogation laser that has numerous narrow-linewidth sidebands. Any adjacent sidebands with equal intensities are separated by the modulation frequency $\nu_{m}$, and each sideband keeps the spectral purity and phase coherence. The total number of spectral lines (NSLs) of the interrogation laser is (2\textit{j}+1), and \textit{j} is the modulation order. Since fully structured interferometry based on ULE glass spacers has been detailedly studied \cite{Judith2019T}, in the proposed scheme such a vibration-tolerable configuration is similarly adopted, and the length of Ramsey free-evolution zone is set to be $\sim$ 9 cm. A magnetic field (typically 5 $\sim$ 7 G) is applied to pick out the field-insensitive $^{3}P_{1}$ (m$_{J}$=0) sublevel.
\\\indent
The spatially separated readout laser corresponding to a fast transition is introduced to measure the population variation of the clock state. For calcium, either a 423 nm transition ($^{1}S_{0}$ -- $^{1}P_{1}$) or 431 nm transition ($^{3}P_{1}$ -- $^{3}P_{0} $) is the appropriate candidate \cite{Shang2017, Judith2019, Hemingway2020}. The polarization of the readout laser is adjusted to be along the \textit{$\hat{x}$} direction to maximize the intensity of the fluorescence, which is collected by the photomultiplier tube. Frequency corrections generated from the readout fluorescence are fed back to the AOM. However, the lineshape of the fluorescence here is different from that of traditional Rabi or Ramsey spectroscopy; the specific characteristics of the two cases are analyzed and discussed below.
\\\indent
For the Rabi excitation, the nonlinear saturation absorption effect under strong fields is considered exclusively for simplicity \cite{Ishikawa1994}, giving an approximate explanation of the velocity-grating Rabi spectroscopy. In the traditional case, the line profile of Rabi spectroscopy for a two-level system follows the approximate expression \cite{Demtroder2008}  
\begin{equation}
L_{Rab}\propto {\rm exp}[-(\frac{\Delta}{0.6\delta\omega_{D}})^{2}]\frac{\gamma/2}{B \sqrt{1-4\Delta^{2}/(A+B)^2}},
\label{E1}
\end{equation}
where $A=\sqrt{\Delta^{2}+\gamma^{2}/4}$, $B=\sqrt{\Delta^{2}+(1+2S)\gamma^{2}/4}$, $\delta\omega_{D}=2\pi\delta\nu_{D}$ is the Doppler width, $\Delta$ the frequency detuning of the OLO to atomic resonance, $\gamma$ the natural linewidth, and $S$ the saturation parameter. The exponential part defines the Gaussian profile of the nonlinear absorption coefficient. The linewidth of the saturation dip $\nu_{s}=\omega_{s}/(2\pi)$ is included in the fractional part, which determines the profile of the saturation dip. \\\indent 
When the OLO with a frequency of $\nu_{L}$ is modulated to be the multi-frequency interrogation laser with frequencies of ($\nu_{L}\pm j\nu_{m}$), the symmetrical sidebands would result in an amplitude-enhanced saturation dip accumulated from the dips at multiple frequencies. This can be concretely explained in terms of the Doppler shift of atoms with specific transverse velocity; when the $\Delta \approx 0$, the atoms with a transverse velocity of $\upsilon_{z}=2\pi p\nu_{m}/k$ (\textit{k} is the wavevector of light, and the integer $p\leq j$) can both absorb the +$p^{\rm th}$ sideband of one interrogation laser and the -$p^{\rm th}$ sideband of the counter-propagating interrogation laser. Therefore, all atoms with velocities of $2\pi p\nu_{m}/k$ can be saturated interrogated simultaneously and hence make accumulated contributions.
\\\indent 
Even if the $\Delta$ is far from zero, pairs of sidebands able to be absorbed by the same transverse velocity group atoms can still be found. It can be noticed that a series of amplitude-enhanced saturation dips will arise when $\Delta=\pm q\omega_{m}/2$ (the \textit{q} is integer). Thus, the calculated profile of the $\pm q^{\rm th}$ amplitude-enhanced saturation dip is 
\begin{gather}
L_{Rab}^{'}\propto \sum_{n=-\alpha}^\alpha {\rm exp}[-(\frac{\Delta+n\omega_{m}}{0.6\delta \omega_{D}})^{2}]\frac{\gamma/2}{B'}\cdot \nonumber \\
 \frac{1}{\sqrt{1-4(\Delta \mp \omega_{m} \cdot q/2)^2/(A'+B')^2}},  
 \label{E2}
\end{gather}
where $ \alpha=(j-q/2)$, $\omega_{m}=2\pi \nu_{m}$, $A'=[(\Delta \mp \omega_{m} \cdot q/2)^{2}+\gamma^{2}/4]^{1/2}$, $B'=[(\Delta \mp \omega_{m} \cdot q/2)^{2}+(1+2S)\gamma^{2}/4]^{1/2}$. The $\alpha$ can be a half-integer here. According to Eq. (\ref{E2}), the normalized amplitude of these dips varied with the $\Delta$ and the NSLs when $\nu_{m}=6\nu_{\rm {s}}$ (the $\nu_{\rm {s}}$ is set as 200 kHz in the theoretical calculations) is calculated and depicted in Fig. \ref{F2}(a). Actually, these dips will overlap when the $\nu_{m}$ is comparable with the $\nu_{s}$, so that the $\vert q\vert =1, 2$ dips will impact the amplitude of the central one. Since just the most-central dip ($q=0$) is concerned, only five central dips are given in Fig. \ref{F2}(a). Specifically, the lineshape of the dip for the cases of NSLs=3 and 251 are shown in the insets of Fig. \ref{F2}(a). From Eq. (\ref{E2}), the amplitude-enhanced ratio $\mathcal{R}$ decreases as the $\nu_{m}$ increases because of fewer atoms contributing to the lineshape; the results are presented in Fig. \ref{F2}(b). Furthermore, the $\mathcal{R}$ would saturate with the increased NSLs since the transverse velocity distribution of the atoms is finite. In the condition of $\delta\nu_{D}$=100 MHz and $\nu_{m}=6\nu_{\rm {s}}$, a maximum $\mathcal{R}$ of 90 can be acquired for the Ca system.
\\\indent
Different from the Rabi excitation, the complicated evolution processes of the atomic state in the optical Ramsey excitation will be explained below with the aid of the pseudo-spin picture. The atomic state denoted by the pseudo-spin vector in the Bloch sphere varies under the action of the ``torque" vector $\vec{\Omega}$. Considering the Doppler effect, the vector can be expressed as $\vec{\Omega}=(-\Omega_{R}, 0, k\upsilon_{z}-\Delta)$, where $\Omega_{R}$ is the Rabi frequency, and the last term of the $\vec{\Omega}$ represents the frequency detuning. In the traditional single-frequency case, for a particular group of atoms with large velocities $\upsilon_{z} \pm \delta \upsilon_{z}/2$ ($\delta \upsilon_{z}<0.3 $ mm/s), the vector $\vec \Omega$ is near the $\widehat{S}_{z}$ axis because $\vert k\upsilon_{z}\vert\gg \Omega_{R}$ (when the $\Delta \approx0$). The pseudo-spin vector initially located at the south pole is rotated after the first interrogation field; see the blue points in Fig. \ref{F3}(a). Experiencing a dark time of $\sim 10^{-4} $ s, the vector fans out since the $k\delta \upsilon_{z}$ is still significant in the optical regime despite a small $\delta \upsilon_{z}$. From the blue points in Fig. \ref{F3}(c), it can be seen that the state vectors still distribute over the southern hemisphere under the excitation of the subsequent $\pi$/2 laser field, showing low transition probabilities for the atoms with large transverse velocity.
\\\indent
But at the case of a multi-frequency interrogation laser with frequencies of ($\nu_{L} \pm j\nu_{m}$), a group of atoms with velocity $\upsilon_{z}$ satisfying $k\upsilon_{z}=2\pi p\nu_{m}$ (the integer $p\leq j$) must exist. For such a group of atoms interacting with one sideband of the multi-frequency laser with a frequency of $(\nu_{L}+p\nu_{m})$, the vector $\vec \Omega=(-\Omega_{R},0,k\upsilon_{z}-(\Delta+2\pi p\nu_{m}))=(-\Omega_{R},0,-\Delta)$, which is the same as that of the zero-transverse-velocity atoms. The evolutions of the atomic state are given for atoms with the velocity of $\upsilon_{z} \pm \delta \upsilon_{z}/2$; see the red points in Figs. \ref{F3}(a)--(c), similar to the evolution processes of the atoms with near-zero transverse velocities \cite{Borde1984}. Therefore, it can be found that almost all atoms with velocity $\upsilon_{z}=2\pi p\nu_{m}/k$ can now be interrogated. Moreover, for the last single pair of interrogation laser fields, the sign of the wavevector is opposite, and then the atoms with velocity $\upsilon_{z}$ can be excited by the sideband with a frequency of $(\nu_{L}-p\nu_{m})$, leading to $\vec \Omega=(-\Omega_{R},0,-\Delta)$ as well. Different from the traditional optical Ramsey method, the proposed method can make the most of the atoms with nonzero transverse velocities producing fringes, and thus the velocity-grating Ramsey fringes can be seen as the superposition of thousands of traditional Ramsey fringes in the frequency domain.
\begin{figure}[t]
\centering
\subfigure{
\includegraphics[width=8cm,height=5.7cm]{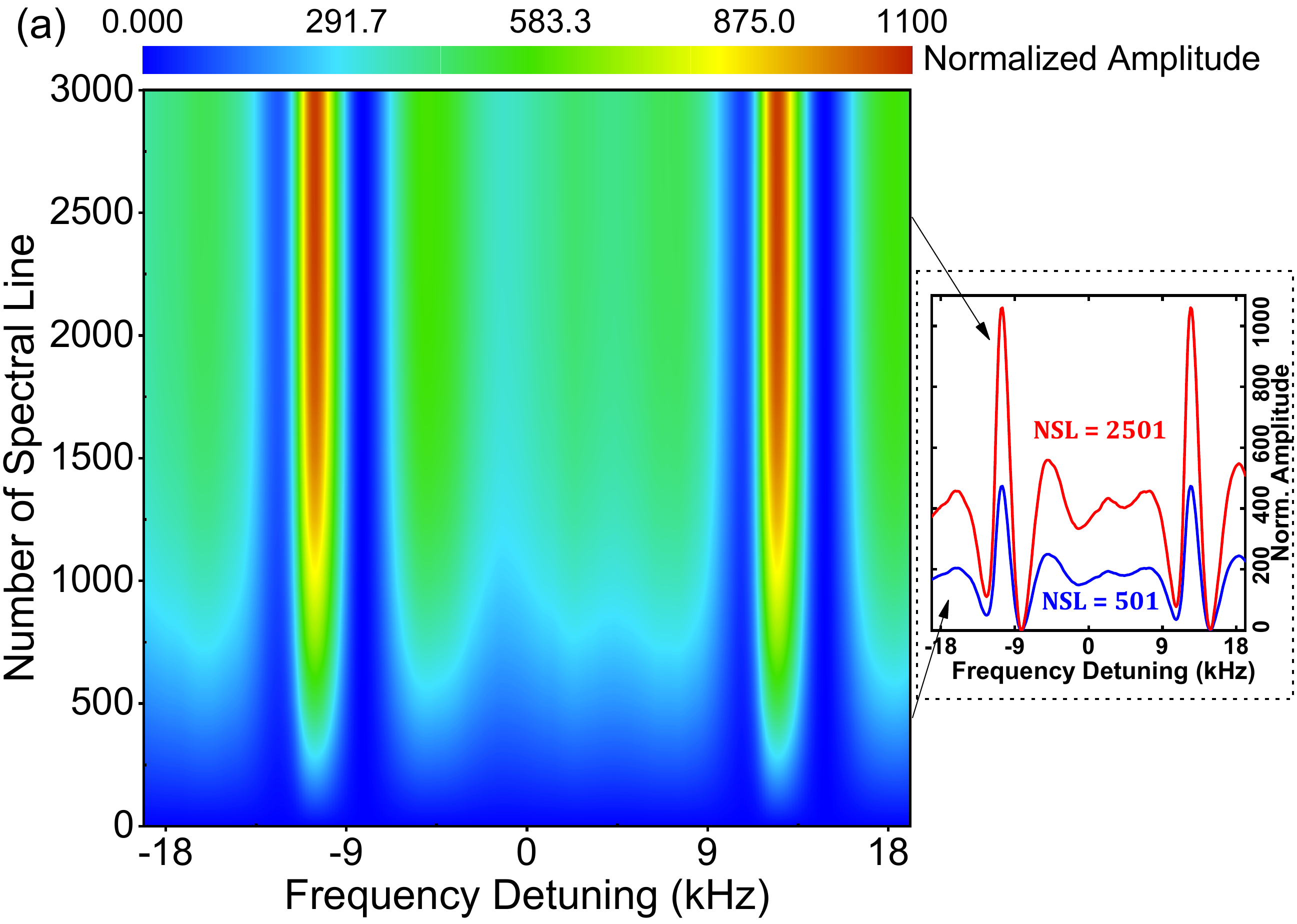}}
\subfigure{
\includegraphics[width=8.3cm,height=3.2cm]{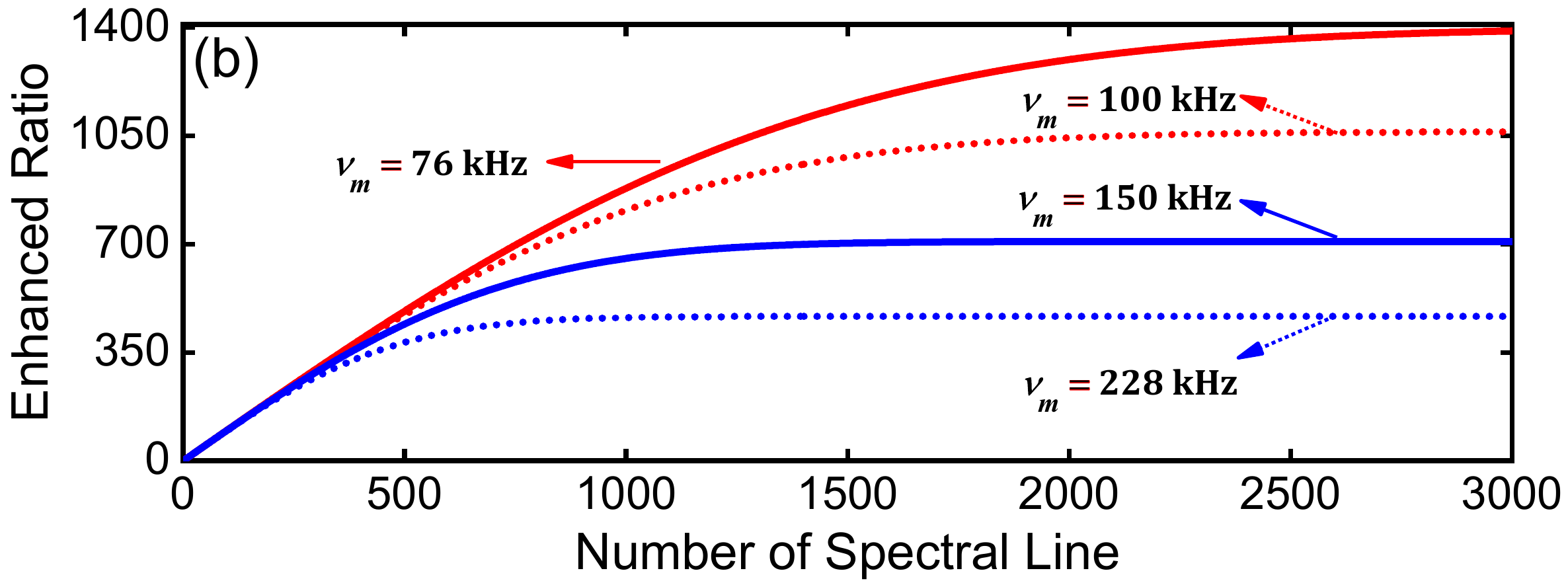}}
\caption{Characteristics of velocity-grating Ramsey fringes. (a) Lineshape of the spectroscopy varied with the $\Delta$ and the NSLs. The insets are the normalized spectroscopy when NSLs=2501 (red line) and NSLs=501 (blue line). (b) Amplitude-enhanced ratio $\mathcal{R}$ of the fringes (one recoil part) vs the NSLs and $\nu_{m}$.} 
\label{F4}
\end{figure}
\\\indent 
Considering the recoil effect but ignoring the relaxation of the state, the transition probability of one atom for the case of single-frequency is given by \cite{Borde1984}
\begin{equation}
P_{e}\propto {\rm Re}\lbrace Ke^{i[2(\Delta \pm \delta)T+\delta \Phi]}\rbrace,
\label{E3}
\end{equation}
where the $K$ is related to the frequency detuning and Rabi frequency, $\delta$ is the photon recoil frequency, $T$ the free-evolution time, and $\delta \Phi$ the phase difference of the four laser fields. The amplitude of the fringes depends on the quantities (or probabilities) of atoms detected at the excited state. Considering that the transverse velocity distribution is Gaussian profile, the amplitudes of the two central Ramsey fringes for the case of multi-frequency are calculated to be
\begin{equation}
L_{Ram}^{'}\propto \sum_{n=-j}^j {\rm exp}[-(\frac{\Delta+n\omega_{m}}{0.6\delta \omega_{D}})^{2}]{\rm Re}\lbrace Ke^{i[2(\Delta \pm \delta)T+\delta \Phi]}\rbrace,
\label{E4}
\end{equation}
where $\delta \omega_{D}$ is the Doppler width. Through numerical calculations, the normalized amplitudes of the Ramsey fringes varied with the NSLs and the $\Delta$ when $\nu_{m}$=100 kHz are given in Fig. \ref{F4}(a). Specifically, the accumulated fringes are shown in the inset of Fig. \ref{F4}(a) when NSLs=501 and 2501. Considering the influence of the $\nu_{m}$, the results of the maximum normalized amplitude of the fringes varied with the $\nu_{m}$ and the NSLs are given as depicted in Fig. \ref{F4}(b). The maximum amplitude-enhanced ratio $\mathcal{R}$ saturates with increasing NSLs and decreases with increasing $\nu_{m}$. It is obvious that a smaller modulation frequency and larger Doppler width are beneficial to improve the $\mathcal{R}$. However, an overly small modulation frequency will disturb the atomic coherence between two adjacent transverse velocity group atoms. The disturbance almost vanishes when the  $\nu_{m}$=228 kHz which is $\sim$ 6 times the contributing span of the Ramsey fringes represented in Eq. (\ref{E3}). When $\nu_{m}$=100 kHz, the maximum $\mathcal{R}$ is 1064. But such a value is unrealistic in the traditional method only relying on the increasing of oven temperature because a nearly 300 K higher temperature is needed. 
\\\indent
For a clock system, the limitations deteriorating the frequency stability typically include QPN and technical noise \cite{Schioppo2017}. Considering that continuous efforts \cite{Ludlow2015, Oelker2019, Matei2017, Zhang2017, Schioppo2017} are aimed at reaching the QPN, the thermal systems are similarly assumed to be not limited by the technical noise eventually. Therefore, specifically for the thermal Ca beam system, due to a substantial improvement of the quantities of the atoms contributing to the clock spectroscopy based on the proposed method, the \textit{S}/\textit{N} of the clock spectroscopy can be enhanced by a factor of $\sqrt{\mathcal{R}}$. Therefore, the frequency stability of the laser would be enormously improved.
\begin{figure}[t]
\centering
\includegraphics[width=7.5cm,height=7.5cm]{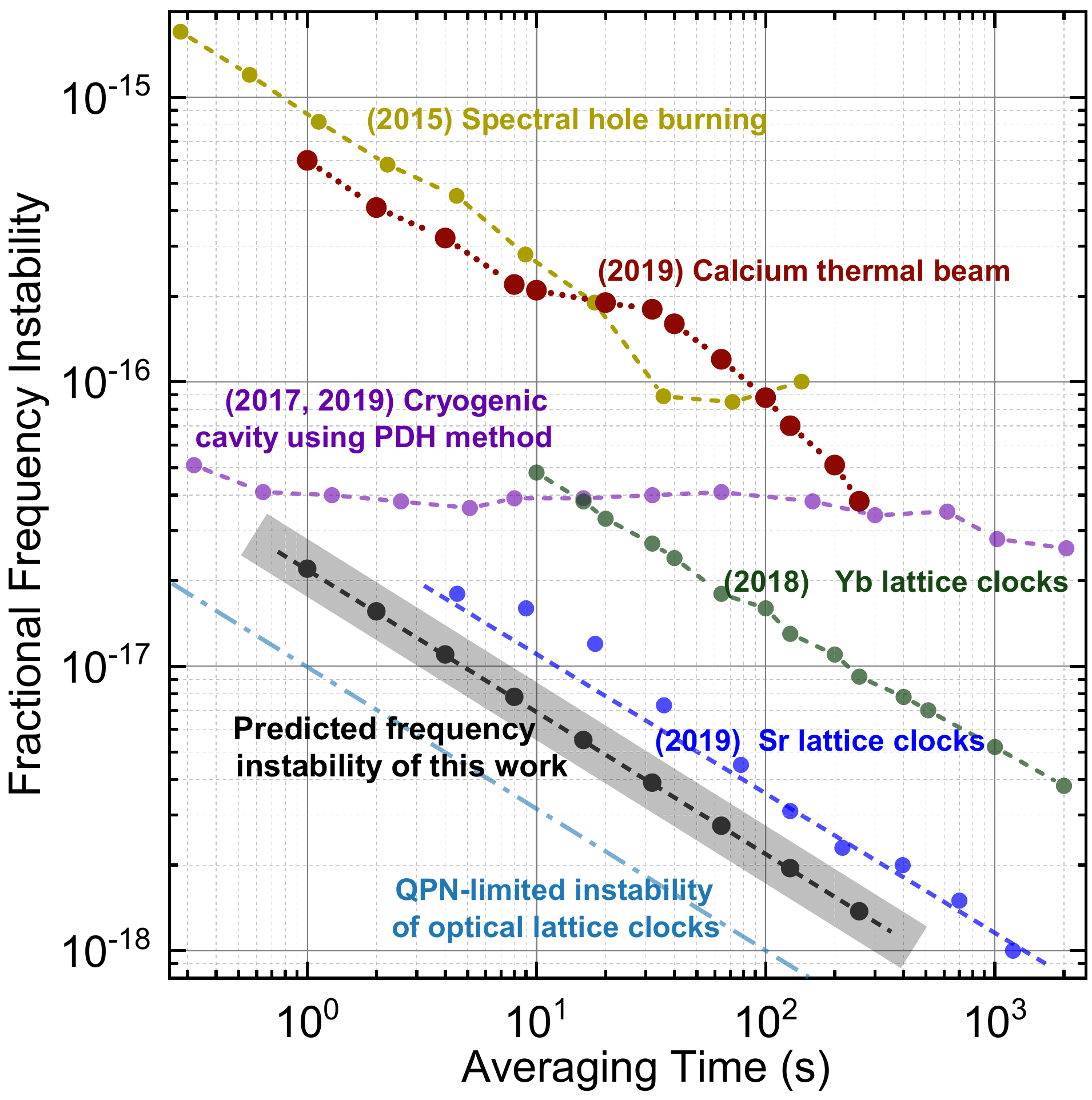}
\caption{Fractional frequency instability of this work compared with other works \cite{Cook2015, Judith2019, Matei2017, Oelker2019, McGrew2018}. The predicted frequency instability of this work is denoted by the gray dashed line with an asymptote of 2.1$\times$10$^{-17}/\sqrt{\tau}$, and the shadow represents the distribution of the frequency instability at different $\nu_{m}$. The light blue dashed line gives one typically unreached QPN-limited frequency instability of optical lattice clocks \cite{Schioppo2017}.} 
\label{F5}
\end{figure}\\\indent
According to the results in Fig. \ref{F4}(b), when the Doppler width is 100 MHz and the linewidth of the Ramsey fringe is $\sim$ 2.0 kHz, the maximum $\mathcal{R}$ of the Ramsey fringes varies from 480 to 1370 at different $\nu_{m}$; thus, the \textit{S}/\textit{N} is inferred to be improved by a factor ranging from 22 to 37, and the frequency instability at 1 s can reach a value between 2.7$\times 10^{-17}$ and 1.6$\times 10^{-17}$ on the basis of the experimental results in \cite{Judith2019, Judith2019T}. Additionally, the frequency instability potentially reach the 10$^{-18}$ level in short timescales. A comparison of the fractional frequency instability of this work with that of other works based on different methods \cite{Cook2015, Judith2019, Matei2017, Oelker2019, McGrew2018, Schioppo2017} is given in Fig. \ref{F5}. It can be seen that the predicted frequency instability of the present work is even comparable with the best-reported atomic clock \cite{Oelker2019} and typical QPN-limited instability of optical lattice clocks \cite{Schioppo2017}. Further improvement on the \textit{S}/\textit{N} can be realized by increasing the atomic flux, e.g., by properly increasing the oven temperature. 
\\\indent
In summary, an ultrastable laser source with $10^{-18}$ frequency instability based on the velocity-grating atomic spectroscopy has been proposed herein, in which numerous nonzero-transverse-velocity atoms can be interrogated and make accumulated contributions to the \textit{S}/\textit{N}. The results show that a QPN-limited fractional frequency instability can be less than 2$\times 10^{-17}/\sqrt{\tau}$, providing a laser source with extraordinary frequency stability. Besides, benefiting from the large feedback bandwidth (on the order of kilohertz \cite{Shang2017, Judith2019, Judith2019T}), the scheme can also provide a method of narrowing the laser linewidth to the mHz level \cite{Robinson2019, Meiser2009, Yu2007}. Such a laser source with superb frequency stability is promising for a wide range of applications, e.g., being used as an optical flywheel in next-generation optical timescales \cite{Milner2019, Judith2019}.
\\
\begin{acknowledgments}
We thank Wei Zhuang for helpful discussions. The work was supported by National Natural Science Foundation of China(NSFC)(91436210) and the National Key Research and Development Program of China.
\end{acknowledgments}


\end{document}